\documentclass[twocolumn,showpacs,amsmath,amssymb]{revtex4}

\usepackage{graphicx}
\usepackage{dcolumn}
\usepackage{bm}

\begin{document}
\title{Three dimensional particle-in-cell simulation of particle acceleration by circularly polarised inertial Alfven waves
in a transversely inhomogeneous plasma}
\author{D. Tsiklauri}
\affiliation{Astronomy Unit, School of Physics and Astronomy, 
Queen Mary University of London, Mile End Road, London, E1 4NS, United Kingdom}
\date{\today}
\begin{abstract}
The process of particle acceleration by left-hand, circularly polarised inertial Alfven waves (IAW)
in a transversely inhomogeneous plasma is studied using 3D particle-in-cell simulation.
A cylindrical tube with, transverse to the background magnetic field, inhomogeneity
scale of the order of ion inertial length is considered on which IAWs with frequency $0.3 \omega_{ci}$ 
are launched that are allowed to develop three wavelength.
As a result time-varying parallel electric fields are generated in the density gradient
regions which accelerate electrons in the parallel to magnetic field direction. Driven perpendicular electric field of IAWs 
also heats ions in the transverse direction. Such numerical setup is relevant for solar flaring loops and
earth auroral zone. This first, 3D, fully-kinetic simulation demonstrates electron
acceleration efficiency in the density inhomogeneity regions, along the magnetic field, of the order of 45\% and 
ion heating, in the transverse to the magnetic field direction,  of 75\%.
The latter is a factor of two times higher than the previous 2.5D analogous study
and is in accordance with solar flare particle acceleration observations.
We find that the generated parallel electric field is localised in the density inhomogeneity region and
rotates in the same direction and with the same angular frequency as the initially launched IAW.
Our numerical simulations seem also to suggest that  the "knee" often found 
in the solar flare electron spectra can alternatively 
be interpreted as the Landau damping (Cerenkov resonance effect) of IAWs
due to the wave-particle interactions.
\end{abstract}	



\maketitle

\section{Introduction}

Super-thermal particles play an important role in many space plasma
situations. The relevant two examples are:
(i) Earths Auroral zone (AZ) that is known to host strong
field-aligned currents, parallel electric field and accelerated particles.
Observations essentially show two modes of particle acceleration present in AZ:
a) Precipitating auroral electrons narrowly peaked at specific energy,
suggestive of a static potential drop  in the AZ (e.g. Ref.\cite{1980SSRv...27..155M});
b) More recent observations by FAST spacecraft (e.g. Ref.\cite{2007JGRA..11205215C}) 
indicate existence of electrons with broad energy and narrow in pitch angle distribution
that is consistent with the inertial Alfven wave (IAW)  acceleration.
(ii) In solar corona, a significant fraction of the energy released
during solar flares is converted into the energy of accelerated
particles \cite{2004JGRA..10910104E}.
The parallel electric field that can accelerate electrons is 
produced when low frequency ($\omega < \omega_{ci}$, where $\omega_{ci}=eB/m_i$ 
is the ion cyclotron frequency) 
dispersive Alfven wave (DAW) has a wavelength perpendicular to the background magnetic field comparable to any of 
the kinetic spatial scales such as: ion gyroradius at electron temperature, 
$\rho_s=\sqrt{k_B T_e/m_i}/ \omega_{ci}$, ion thermal 
gyroradius, $\rho_i=\sqrt{k_B T_i/m_i}/ \omega_{ci}$, \cite{1976JGR....81.5083H} or to electron inertial length 
$\lambda_e = c/ \omega_{pe}$ \cite{1979JGR....84.7239G}. 
Dispersive Alfven waves are divided into Inertial Alfven Waves or Kinetic Alfven Waves (KAW) 
depending on the relation between the plasma $\beta$ and  electron/ion mass ratio $m_e/m_i$ \cite{2000SSRv...92..423S}.
When $\beta \ll m_e/m_i$ (i.e. when Alfven speed is much greater than electron and ion thermal speeds, 
$V_A \gg v_{th,i}, v_{th,e}$) dominant mechanism for sustaining $E_\parallel$ is the parallel electron inertia 
and such waves are called Inertial Alfven Waves. When $\beta \gg m_e/m_i$, (i.e. when  $V_A \ll v_{th,i}, v_{th,e}$) 
then thermal effects become important and the dominant mechanism for 
supporting  $E_\parallel$ is the parallel electron pressure gradient. Such waves 
are called  Kinetic Alfven Waves.  

The context of this study is related to the theoretical plasma physics processes operating
in the particle acceleration in solar flares. Ref.\cite{2009A&A...508..993B}'s Introduction gives
a good overview of both theoretical and observational unresolved issues. If the solar flare particle
acceleration happens in the corona, in order to explain the observed X-ray flux as the
electrons smash into the dense layers of the Sun, implausibly large acceleration region volumes
and/or densities are required. Different ideas have been put forward: 
(i) Substantial re-acceleration in the chromosphere
of electrons accelerated in and injected from the corona 
can greatly reduce the density and number of fast electrons
needed to produce a X-ray flux \cite{2009A&A...508..993B};  However, this still
seems problematic because recent analysis of the evolution of a radio spectrum from
a dense flare \cite{2007ApJ...666.1256B} (Bastian et al. 2007) shows that a significant fraction of the energy in the energetic
electrons can be deposited into the coronal loop, as opposed to into the chromosphere. 
The estimates show that the energy deposited in the corona can approach about 30\% flare energy.
(ii) The electric circuit formed by precipitating and returning electrons \cite{2011SSRv..159..357Z,2011A&A...532A..17Z}.
The effect was found from the Ampere law which handles the circuit of injected and 
returning electrons when simulated  with Particle-In-Cell and Fokker-Planck simulations.
(iii) When wave-particle interactions in non-uniform plasma are taken into account, 
the evolution of the Langmuir wave spectrum towards smaller wave-numbers,  
leads to an effective acceleration of electrons. Thus, the time-integrated spectrum of non-thermal electrons 
shows an increase in super-thermal electrons, because of their acceleration by the 
Langmuir waves \cite{2012A&A...539A..43K}.
(iv) Solar flare triggered DAWs, as opposed to the electron beams, propagating towards
the solar coronal loop foot-points, and accelerating electrons along the propagation paths \cite{2008ApJ...675.1645F}.
Key ingredient in such approach is the transverse density inhomogeneity (i.e. edges of solar coronal
loops), which enable DAW to acquire non-zero perpendicular to the magnetic field wavelength, comparable
to the above mentioned kinetic scales. This results in the generation of parallel electric fields, that
can effectively accelerate electrons \cite{2005A&A...435.1105T,2008PhPl...15k2902T,2011PhPl...18i2903T}.

Introduction section of Ref.\cite{2011PhPl...18i2903T} gives an overview of the previous work on this topic
in some detail. Here we mention a latest addition, Ref.\cite{2011JGRA..11600K15M}, which
studies the interaction of an isolated Alfven wave packet with a plasma density cavity.
Ref.\cite{2011PhPl...18i2903T} considered particle acceleration by DAWs in the 
transversely inhomogeneous plasma via full kinetic simulation particularly focusing on
the effect of polarisation of the waves and different regimes (inertial and kinetic).
In particular, Ref.\cite{2011PhPl...18i2903T}  studied particle 
acceleration by the low frequency ($\omega=0.3\omega_{ci}$)
DAWs, similar to considered in Ref.\cite{2005A&A...435.1105T,2008PhPl...15k2902T}, in {\it 2.5D geometry},     
focusing on the effect of the wave polarisation, left- (ion cyclotron branch) and right- (whistler branch) 
circular and elliptical,
in the different regimes inertial ($\beta < m_e/m_i$) and kinetic 
($\beta > m_e/m_i$). A number of important conclusions were drawn, including
(i) The fraction of accelerated electrons (along the magnetic field), in
the density gradient regions is 20\%-35\% in 2.5D geometry.
(ii) While keeping the power of injected DAWs the same in all
considered numerical simulation runs, in the case of right circular, left
and right elliptical polarisation DAWs with $E_y/E_z=6$ (with $x$ being the direction
of the uniform background magnetic field) produce more
pronounced parallel electron beams.
(iii) The parallel electric field for solar flaring plasma parameters
exceeds Dreicer electric field by eight orders of magnitude.
(iv) Electron beam velocity has the phase velocity of the DAW. This
can be understood by Landau damping of DAWs. The mechanism can
readily provide electrons with few tens of keV.
(v) When in 2.5D case the mass ratio was increased from $m_i/m_e=16$ to
73.44,  the fraction of accelerated electrons has increased from 20\% to
30-35\% (depending on DAW polarisation).
This is because the velocity of the beam has shifted to lower velocity.
As there are always more electrons with a smaller velocity than
higher velocity in the Maxwellian distribution, one can conjecture that for the mass ratio
$m_i/m_e=1836$ the fraction of accelerated electrons would be even higher
than 35\%.

In the present work we focus on the 3D effects on particle acceleration and parallel electric field generation.
In particular, instead of 1D transverse, to the magnetic field, density (and temperature) inhomogeneity,
we consider the 2D transverse density (and temperature) inhomogeneity in a form of a circular cross-section
cylinder, in which density (and temperature) varies smoothly across the uniform magnetic field
that fills entire simulation domain. Such structure mimics a solar coronal loop which is kept in 
total pressure balance.
Section II describes the model for the numerical simulation, while the results are presented in section III.
We close the paper with the conclusions in section IV. 

\section{The model}

The general observational context of this work in outlined in Fig.~1, which shows that a solar flare
at the solar coronal loop apex triggers high frequency, $0.3 \omega_{ci}$, DAWs which then propagate 
towards loop footpoints. 
We only consider a single DAW harmonic, however in reality flare may trigger
a wide spectrum of waves with different wave-numbers and frequencies, prescribed by a
turbulent cascade. Some examples (albeit longitudinal (Langmuir), not transverse (Alfvenic) type) of creating of such 
turbulence can be found in Ref.\cite{2011ApJ...733...33Z}. We conjecture that if the electron beam 
has non-zero velocity component, transverse to the background magnetic field, 
kinetic plasma instabilities, such as electron cyclotron maser will also generate
the transverse (Alfvenic) turbulent cascade.
Another issue to consider in the context of DAW excitation is the possibility of
excitation of KAWs or IAWs by means of magnetic field-aligned currents -- essentially electron beams
drifting with respect to stationary ions \cite{2012ApJ...754..123C} or fast ion beam excitation \cite{1996SoPh..168..219V}. Ref.\cite{2011PhRvE..84d6406C}
considered the situation when KAWs are excited by current (fluid) instability. 
The instability condition for this excitation by current is satisfied 
even for small drift velocity, e.g $v_D=0.1v_A$, when KAWs can effectively grow.
However, Ref.\cite{2011PhRvE..84d6406C} 
did not include the resonant excitation of DAWs by the inverse Landau damping 
because its instability condition requires a larger drift velocity, in general, 
larger than the Alfven velocity. In a different regime, 
the importance of the Landau (Cerenkov) resonance for
the particle acceleration and parallel electric field generation by the DAWs was stressed by
Refs.\cite{2011PhPl...18i2903T,2011A&A...527A.130B}.

In our model (see Fig.~1) the transverse density (and temperature) inhomogeneity scale is of the order of
$30$ Debye length ($\lambda_D$) that for the considered mass ratio $m_i/m_e=16$ corresponds to 0.75 
ion inertial length $c/\omega_{pi}$. Existence of such thin loop threads, tens of cm wide, in the solar corona 
is of course open to debate. The finest loops observed with TRACE have a width of the
order of 1000-2000 km and have a monolithic structure with 
a single temperature at any cross-section, thus presently not supporting the  
thin thread concept. The TRACE CCD camera
has 0.5 arcsec pixels which is 366 km on the sun. Therefore it 
is plausible that the smallest observed 1000 km wide monolithic
structures are probably too close to the resolution limit. The future
high spatial resolution space missions such as Solar Probe Plus may possibly reveal the loop sub-structuring.

We use EPOCH (Extendable Open PIC Collaboration) a multi-dimensional, fully electromagnetic, 
relativistic particle-in-cell code which was developed and is used by 
Engineering and Physical Sciences Research Council (EPSRC)-funded 
Collaborative Computational Plasma Physics  (CCPP) consortium of 30 UK researchers.
We use 3D version of the EPOCH code. 
The relativistic equations of motion are solved
for each individual plasma particle.
The code also solves Maxwell's equations, with self-consistent currents, using 
the full component set of EM fields
$E_x,E_y,E_z$  and $B_x,B_y,B_z$.
EPOCH uses SI units. For the graphical
presentation of the results, here we use the following normalisation:
Distance and time are normalised to $c / \omega_{pe}$ and $\omega_{pe}^{-1}$, while
electric and magnetic fields to $\omega_{pe}c m_e /e$ and  $\omega_{pe} m_e /e$ respectively.
When visualising the normalised results we use   $n_0 =10^{16}$ m$^{-3}$ in the
least dense parts of the domain ($y=0$, $y=y_{m}$, $z=0$, $z=z_{m}$), which are located at the edges of the
simulation domain (i.e. fix $\omega_{pe} = 5.64 \times 10^9$ Hz radian on the domain edges).
Here $\omega_{pe} = \sqrt{n_e e^2/(\varepsilon_0 m_e)}$ is the electron plasma frequency,
$n_\alpha$ is the number density of species $\alpha$ and all other symbols have their usual meaning.
The spatial dimension of the simulation box is fixed
at $x=5000$ and $y=z=200$ grid points for the mass ratio $m_i/m_e=16$.
This mass ratio value corresponds to the in the inertial Alfven wave (IAW) regime
because plasma beta in this study is fixed at 
$\beta= 2 (v_{th,i}/c)^2(\omega_{pi}/\omega_{ci})^2 = 
n_0(0,0)k_B T /(B_0^2/(2\mu_0))=0.02$. Thus $\beta=0.02 < m_e/m_i=1/16=0.0625$;
This is the maximal value that can be considered with the
available computational resources. Using 720 processor cores, the presented in this paper numerical run 
took 9 days and 14 hours.
The grid unit size is  $\lambda_D$. 
Here $\lambda_D = v_{th,e}/ \omega_{pe}=5.345\times 10^{-3}$ m
is the Debye length ($v_{th,e}=\sqrt{k_B T/m_e}$ is electron thermal speed).
This makes the spatial simulation domain size of $x=[0,x_{m}]=[0,26.727m]$, $y=[0,y_{m}]=[0,1.069m]$
$z=[0,z_{m}]=[0,1.069m]$. Particle velocity space is resolved  (i.e distribution functions produced in $V_x, V_y, V_z$ directions) with
30000 grid points with particle momenta in the range $\mp 6.593\times10^{-22}$ kg m s$^{-1}$. 
We use $2.4\times 10^9$ electrons and the same number of protons in the simulation.
In principle, plasma number density $n(y,z)$ (and hence $\omega_{pe}(y,z)$) can be regarded 
as arbitrary, because we use $\omega_{pe}(0,0)$ in our normalisations.
We impose constant background magnetic field $B_{0x} =320.753$ Gauss along
$x$-axis. This corresponds to $B_{0x}=1.0 (\omega_{pe} m_e /e)$, so that
with the normalisation used for the visualisation
purposes, normalised background magnetic field is unity. 
This sets $\omega_{ce}/\omega_{pe} = 1.00$.
Electron and ion temperature at the simulation box edge
is also fixed at $T(0,0,0)=T_e(0,0,0)=T_i(0,0,0)=6\times10^7$K. 
This in conjunction with $n_0(0,0) =10^{16}$ m$^{-3}$
makes plasma parameters similar to that of a dense flaring loops
in the solar corona.

\begin{figure}[htbp]    
\centerline{\includegraphics[width=0.49\textwidth]{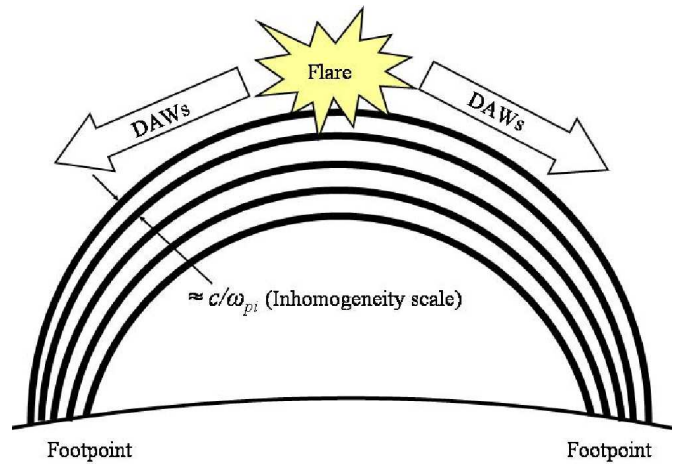}}
\caption{A conceptual sketch of the model. A solar flare launches DAWs from the 
solar coronal loop apex which rush down towards the photospheric 
footpoints, accelerating electrons along the propagation direction and heating ions in the transverse 
direction. The transverse inhomogeneity scale is $\approx 30$ Debye length ($\lambda_D$) that 
for the considered mass ratio $m_i/m_e=16$ corresponds to $0.75 c/\omega_{pi}$.}
   \end{figure}

\begin{figure}[htbp]    
\centerline{\includegraphics[width=0.49\textwidth]{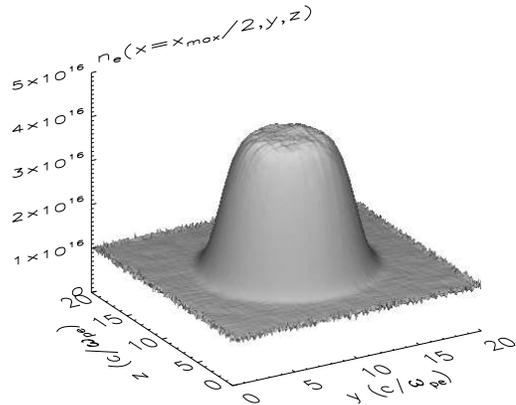}}
\caption{Shaded surface plot of $n_e(x=x_{m}/2,y,z)$ at $t=0$.
}
   \end{figure}

\begin{figure}[htbp]    
\centerline{\includegraphics[width=0.49\textwidth]{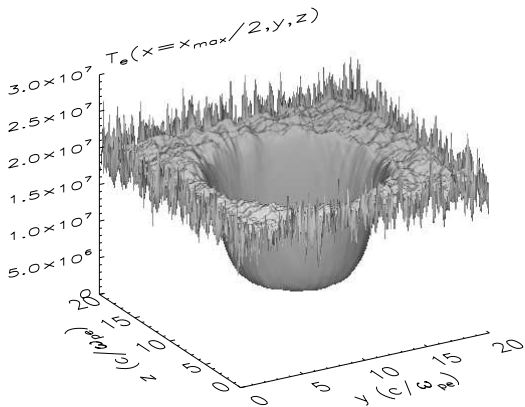}}
\caption{Shaded surface plot of $T_e(x=x_{m}/2,y,z)\propto 1/n_e(x=x_{m}/2,y,z)$ at $t=0$.
 }
   \end{figure}

\begin{figure*}[htbp]    
\centerline{\includegraphics[width=0.8\textwidth]{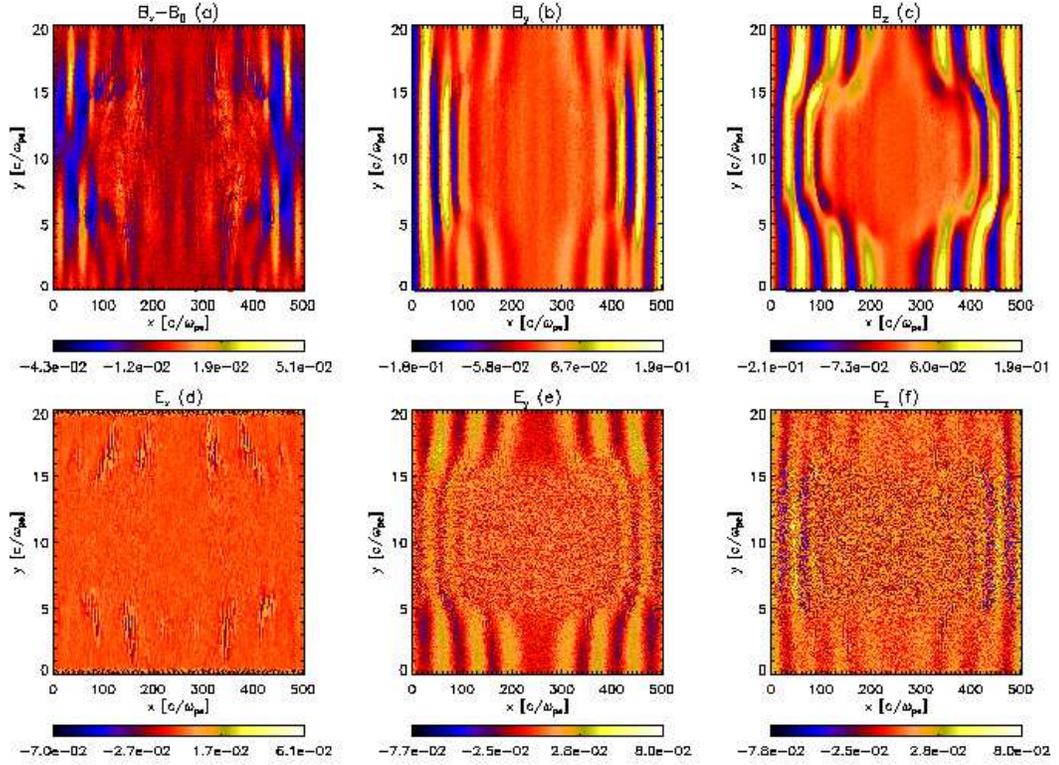}}
\caption{Contour (intensity) plots of the following physical quantities at $t_{end}=75\omega_{ci}^{-1}$:
(a) $B_x(x,y,z=z_{m}/2)-B_0$,
(b) $B_y(x,y,z=z_{m}/2)$, (c) $B_z(x,y,z=z_{m}/2)$,
(d) $E_x(x,y,z=z_{m}/2)$, (e) $E_y(x,y,z=z_{m}/2)$, (f) $E_z(x,y,z=z_{m}/2)$. These correspond
to a vertical cut through the mid-plane $z=z_{m}/2$.
}
\end{figure*}

\begin{figure*}[htbp]    
\centerline{\includegraphics[width=0.8\textwidth]{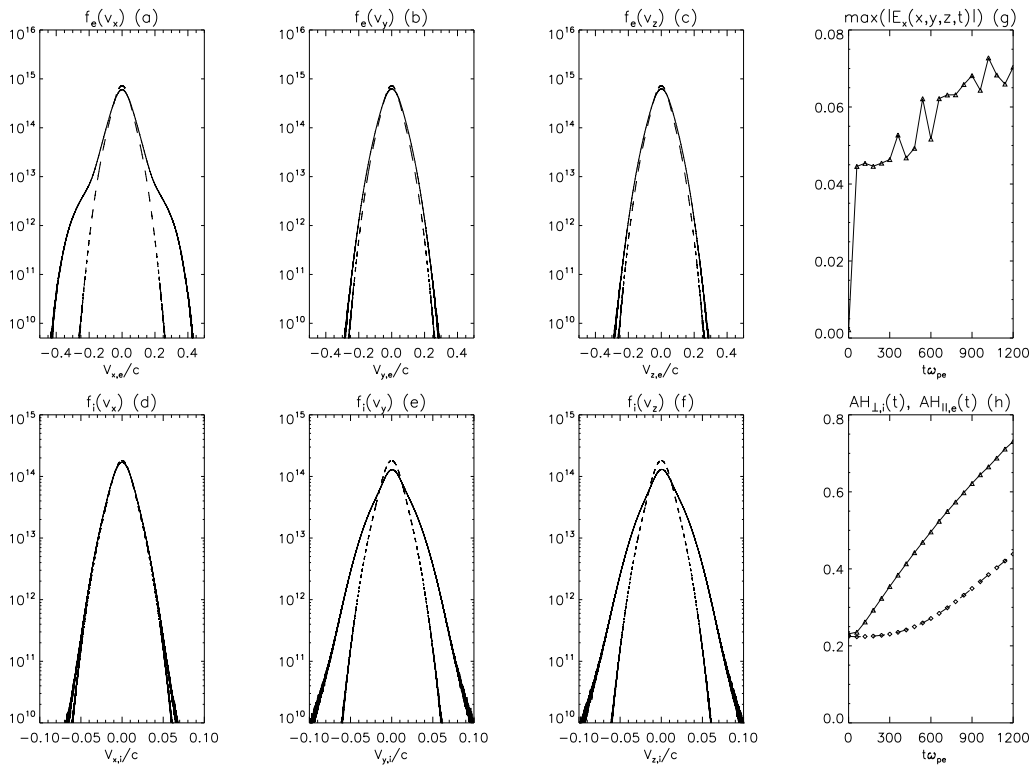}}
\caption{Time evolution (at $t=0$ and $t=t_{end}=75\omega_{ci}^{-1}$) of electron and ion velocity distribution 
functions versus velocity $x,y$ and $z$ components on a log-linear plot: 
(a) $f_e(v_x,t=0)$ dashed (inner) curve and $f_e(v_x,t=t_{end})$ solid (outer) curve,
(b) $f_e(v_y,t=0)$ dashed (inner) curve and $f_e(v_y,t=t_{end})$ solid (outer) curve,
(c) $f_e(v_z,t=0)$ dashed (inner) curve and $f_e(v_z,t=t_{end})$ solid (outer) curve, 
(d) $f_i(v_x,t=0)$ dashed (inner) curve and $f_i(v_x,t=t_{end})$ solid (outer) curve,
(e) $f_i(v_y,t=0)$ dashed (inner) curve and $f_i(v_y,t=t_{end})$ solid (outer) curve,
(f) $f_i(v_z,t=0)$ dashed (inner) curve and $f_i(v_z,t=t_{end})$ solid (outer) curve.
Time evolution (at 20 time intervals between $t=0$ and $t=t_{end}$) of the
following: (g) $\max |E_x(x,y,z,t)|$, triangles connected with a solid curve and
(h) $AH_{\parallel,e}(t)$ index, diamonds connected with dashed curve, according to
Eq.(7), 
$AH_{\perp,i}(t)$ index, triangles connected with a solid curve, according to
Eq.(8).}
\end{figure*}

\begin{figure*}[htbp]    
\centerline{\includegraphics[width=0.8\textwidth]{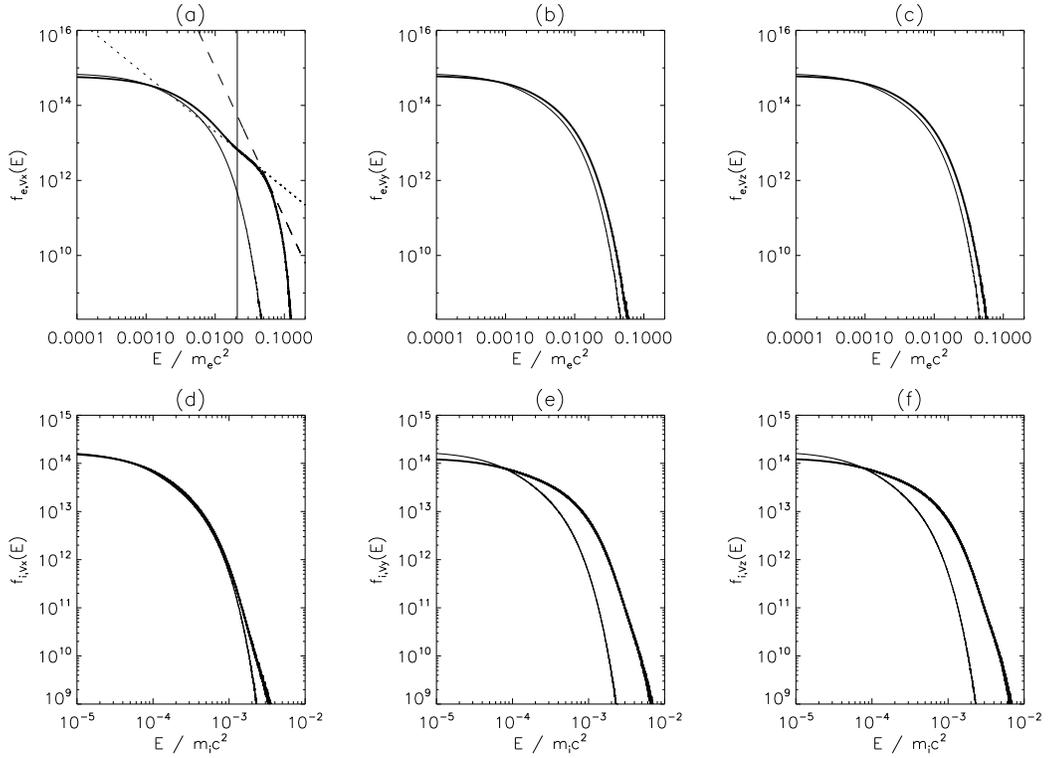}}
\caption{Time evolution (at $t=0$ and $t=t_{end}=75\omega_{ci}^{-1}$) of electron and ion velocity distribution 
functions (with positive velocities) versus kinetic energy of particles $E$ on a log-log plot.
Top row: 
(a) $f_{e,Vx}(E,t=0)$ thin solid (inner) curve and $f_{e,Vx}(E,t=t_{end})$ thick solid (outer) curve,
(b) $f_{e,Vy}(E,t=0)$ thin solid (inner) curve and $f_{e,Vy}(E,t=t_{end})$ thick solid (outer) curve, 
(c) $f_{e,Vz}(E,t=0)$ thin solid (inner) curve and $f_{e,Vz}(E,t=t_{end})$ thick solid (outer) curve. 
Vertical line in (a) corresponds to the kinetic energy, $E_{DAW}= (\gamma-1) m_e c^2$, 
with $\gamma=1/\sqrt{1-0.201^2}$ and $0.201c$ being phase velocity of left-hand polarised IAW in the low
density region.
Dotted line is $2\times10^{10}E^{-1.5}$ and dashed line is $10^7E^{-4}$.
Bottom row: 
(d) $f_{i,Vx}(E,t=0)$ thin solid (inner) curve and $f_{i,Vx}(E,t=t_{end})$ thick solid (outer) curve,
(e) $f_{i,Vy}(E,t=0)$ thin solid (inner) curve and $f_{i,Vy}(E,t=t_{end})$ thick solid (outer) curve, 
(f) $f_{i,Vz}(E,t=0)$ thin solid (inner) curve and $f_{i,Vz}(E,t=t_{end})$ thick solid (outer) curve. 
}
\end{figure*}

\begin{figure*}[htbp]    
\centerline{\includegraphics[width=0.8\textwidth]{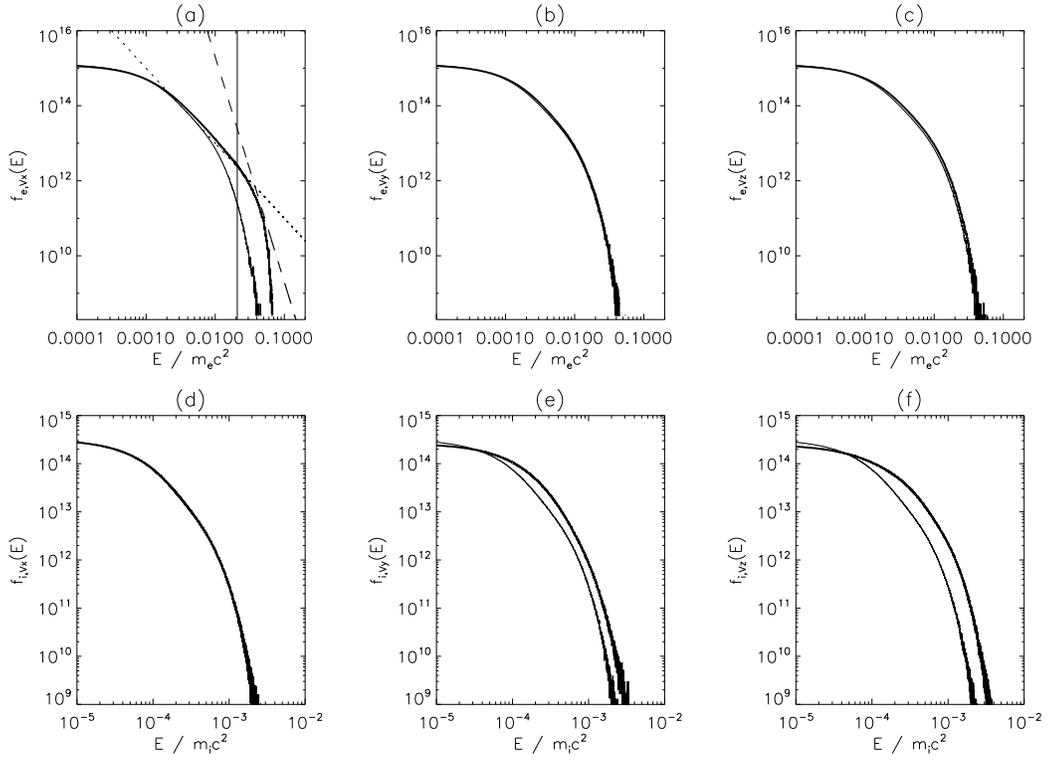}}
\caption{
As in Fig.~6 but for run L16 from Ref.\cite{2011PhPl...18i2903T}.
Dotted line is $10^{9}E^{-2}$ and dashed line is $2\times10^3E^{-6}$.}
\end{figure*}

\begin{figure*}[htbp]    
\centerline{\includegraphics[width=0.8\textwidth]{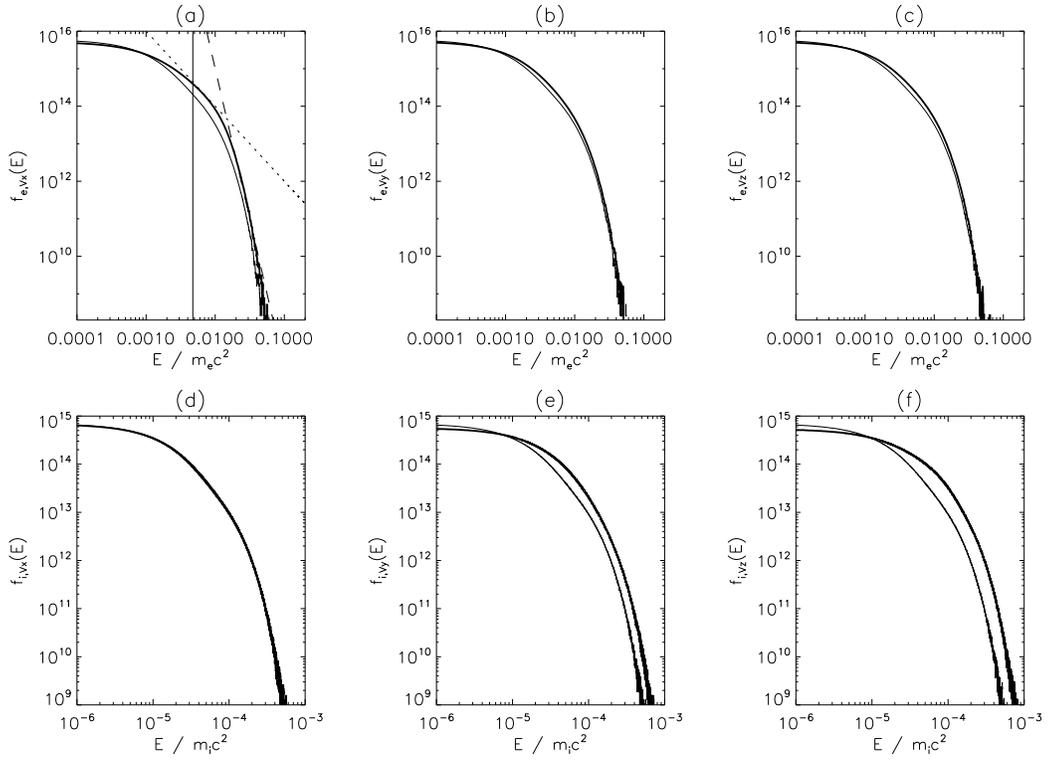}}
\caption{
As in Fig.~6 but for run L73 from Ref.\cite{2011PhPl...18i2903T}.
Vertical line in (a) corresponds to the kinetic energy, $E_{DAW}= (\gamma-1) m_e c^2$, 
with $\gamma=1/\sqrt{1-0.097^2}$ and $0.097c$ being phase velocity of left-hand polarised IAW in the low
density region.
Dotted line is $10^{10}E^{-2}$ and dashed line is $10^{-1}E^{-8}$.
}
\end{figure*}

We consider a transverse to the background magnetic field 
variation of number density as following
\begin{eqnarray}
{n_i(y)}= {n_e(y)}=&   \nonumber \\
1+3 \exp\Biggl[- 
\left(\frac{   \sqrt{(y-y_0)^2+(z-z_0)^2} }{50 \lambda_D}\right)^6\Biggl] =&   \nonumber \\
1+3 \exp\Biggl[- \left(\frac{|\vec r-\vec r_0|}{50 \lambda_D}\right)^6\Biggl]
\equiv f(y,z).&
\end{eqnarray}
Eq.(1) implies that in the central region (across the 
$y$ and $z$ directions), the density is
smoothly enhanced by a factor of 4, and there are the 
strongest density gradients having 
a width of about ${30 \lambda_D}$ around the 
points $r_s=\sqrt{y^2+z^2}=51.5 \lambda_D$ and $r_s=148.5 \lambda_D$.
Here $r_0=(y_0,z_0)=(100\lambda_D,100\lambda_D)$.
Fig.~2 shows $n_e(x=x_{m}/2,y,z)$ at $t=0$, which indicates that
the number density at the cylindrical tube centre increases by factor of 4
compared to the exterior. This density behaviour represents the solar coronal loop.
The background temperature of ions and electrons are varied accordingly
\begin{equation}
{T_i(y,z)}/{T_0}=
{T_e(y,z)}/{T_0}=f(y,z)^{-1},
\end{equation}
such that the thermal pressure remains constant. Because the background magnetic field
along the $x$-coordinate  is constant, the total pressure is also constant. 
Fig.~3 shows $T_e(x=x_{m}/2,y,z)\propto 1/n_e(x=x_{m}/2,y,z)$ at $t=0$.
Note how the temperature at the cylindrical tube centre decreases by factor of 4
compared to the exterior.

The DAW is launched by driving domain left edge, $x=1\lambda_D$, as follows
\begin{eqnarray}
E_y(1,y,z,t+\Delta t)= E_y(1,y,z,t)- &   \nonumber \\
A_y\sin(\omega_d t)\left(1-\exp\left[-(t/t_0)^2\right]\right),& 
\end{eqnarray}
\begin{eqnarray}
E_z(1,y,z,t+\Delta t)=E_z(1,y,z,t)- &   \nonumber \\
A_z\cos(\omega_d t)\left(1-\exp\left[-(t/t_0)^2\right]\right),&
\end{eqnarray}
which produces the case of left-polarised DAW, i.e. ion cyclotron wave.
$\omega_d$ denotes the driving frequency which was fixed at $\omega_d=0.3\omega_{ci}$.
The fact that $\omega_d < \omega_{ci}$   ensures that no ion-cyclotron damping is present and also
that the generated DAW is well on the Alfven wave branch with dispersion
properties similar to the Alfven wave (cf. Fig. 3.4 from Ref.\cite{dendy}). 
$t_0$ is the onset time of the driver, which was fixed at $3.0 \omega_{ci}^{-1}$
i.e.  $48.000  \omega_{pe}^{-1}$ for the case of $m_i/m_e=16$. 
Thus the driver onset time is $9 \omega_{ci}^{-1}$ (9 comes from $3^2$). 
The initial amplitudes of the $E_\perp$ are chosen 
as $A_y=A_z=0.05(cB_{0x})$. Therefore magnetic perturbation amplitude of DAW is 5\% of the background magnetic field,
thus the simulation is weakly non-linear.
Such transverse electric field $E_\perp=E_{y,z}$ driving generates 
L-circularly polarised DAW. For further details on polarisation see  paragraph before 
Eq.(7) from Ref. \cite{2011PhPl...18i2903T}.
In the case of L-circularly polarised DAW the dispersion is
\begin{equation}
k=\frac{\omega}{c}\left(1+\frac{\omega_{pe}^2+\omega_{pi}^2}{(\omega_{ce} + \omega)
(\omega_{ci} - \omega)}\right)^{1/2}.
\end{equation}
Eq.(5) shows that when $\omega=\omega_{ci}$ the ion-cyclotron resonance occurs.

\section{Results}

Fig.~4 presents the numerical simulation results for the electromagnetic fields. We see from
Figs.4(e) and 4(f) that in response to the
transverse electric field driving, which proceeds according to Eqs.(3) and (4),  left-hand, circularly polarised IAW
has developed three wavelength. If we use Eq.(5), then
for the considered plasma parameters, $3 \times (2 \pi / k) / (c/\omega_{pe}) = 201.7$. We gather from Fig.~4(e)
that the waveform at $y=0$ has travelled to $x \approx 200$. It has not reached $z=201.7$ quite yet because the  driving 
onset time is $9 \omega_{ci}^{-1}$. We see that $E_y$ component develops phase mixing i.e. waveform distorts across the
transverse $y$ and $z$ coordinates generating strong transverse gradients. The transverse magnetic
fields have similar behaviour to transverse electric fields. However, phase-mixed field component is $B_z$ (Fig.~4(c)).
Movie that shows time dynamics of $B_z(x,y,z=z_{m}/2)$ (a plane that cuts through the middle of the over-dense
tube, along the background magnetic field) is
shown in the supplementary material \cite{movies}.
Fig.~4(d) shows that parallel electric field is generated only in the density inhomogeneity regions $4 < r < 7$ and
$13<r<16$, where $r=\sqrt{y^2+z^2}$. The time evolution of this parallel electric field is shown
in two different cross-sections in the supplementary material \cite{movies}:   $E_x(x,y,z=z_{m}/2)$
(a plane that cuts through the middle of the over-dense
tube, {\it along} the background magnetic field) and  $E_x(x=x_{m}/8,y,z)$
(a plane that cuts through the 1/8th of the over-dense tube, {\it across} the background magnetic field).
We gather from movie 2 \cite{movies} that in this 3D simulation, in the plane cut along the background magnetic field,
behaviour is similar to 2.5D results both in PIC simulation (see e.g. Fig.~1 from Ref.\cite{2011PhPl...18i2903T})
or cold plasma approximation (see animation 2 (in Supplementary Data) from Ref.\cite{2007NJPh....9..262T}).
What is new in 3D is the ability to study the {\it transverse} structure of the parallel electric field.
Somewhat unexpected results are shown in movie 3 \cite{movies}.
The bright yellow blob that can be seen rotating around the centre is a marker to guide ones
eye. It rotates in the clockwise direction with the angular frequency of $0.3\omega_{ci}$.
Its dynamics is described by the equation
\begin{eqnarray}
marker(y,z,t)=2.0\times10^{-2}\exp \Bigl( & \nonumber \\
-\bigl\{[(y-y_{m}/2)\cos(0.3\omega_{ci}t+\pi)- & \nonumber \\
(z-z_{m}/2)\sin(0.3\omega_{ci}t+\pi)-y_{m}/4]/0.5\bigr\}^2 & \nonumber \\
-\bigl\{[(z-z_{m}/2)\cos(0.3\omega_{ci}t+\pi)+ & \nonumber \\
(y-y_{m}/2)\sin(0.3\omega_{ci}t+\pi)-z_{m}/4]/0.5\bigr\}^2 \Bigr).& 
\end{eqnarray}
We see only the marker blob rotating in the clockwise direction until about time
$t \leq 600 \omega_{pe}$. Then the generated parallel electric field reaches the plane corresponding to 
$x=x_{m}/8$ in which this movie is produced. From $t \geq 600 \omega_{pe}$ until the
end of the simulation we see that parallel electric field is localised within circular bands $4 < r < 7$ and
$13<r<16$ and rotating  in the clockwise direction with the frequency  $0.3\omega_{ci}$.
This seems initially surprising why the parallel electric field should rotate.
However, if one recalls that the IAW also rotates in the same clockwise direction, then
this seems logical (that both parts of the wave $E_\perp$ that is used as a driver and
the generated $E_\parallel$ both rotate in the same sense and with same angular frequency).
Note that rotation in movie 3 and translatory motion in movies 1 and 2 \cite{movies} appear
step-like because there are only 20 time snapshots taken throughout the simulation that ends
at $t_{end}=75\omega_{ci}^{-1}=1200\omega_{pe}^{-1}$. Thus in this animation $\Delta t =3.75 \omega_{ci}^{-1}$.
Also, in all movies distances on $x-y$ axes are quoted in number of grid points --
not in $c/\omega_{pe}$ as in the rest of the paper. Thus, the conclusion here is
that the generated parallel electric field is localised in the density inhomogeneity region and
rotates in the same direction and with the same angular frequency as the initially launched DAW.

We quantify the  particle acceleration and plasma heating 
by defining the following indexes:
\begin{eqnarray}
AH_{\parallel,e}(t)=& \nonumber \\
\frac{\int_{|v_{x}| > \langle v_{th,e}\rangle}^\infty      
f_e(v_x,t)dv_x\bigl/\left(x_{m}\pi[(r_{o}-r_0)^2-(r_{i}-r_0)^2]\right)}
 { \int_{-\infty}^\infty f_e(v_x,0) dv_x \bigl/\left( x_{m}y_{m}z_{m}\right)},
\end{eqnarray}
\begin{eqnarray}
AH_{\perp,i}(t)=& \nonumber \\
\frac{\int_{|v_{\perp}| > \langle v_{th,i}\rangle}^\infty      
f_i(v_\perp,t)dv_\perp\bigl/\left(x_{m}\pi[(r_{o}-r_0)^2-(r_{i}-r_0)^2]\right)}
 { \int_{-\infty}^\infty f_i(v_\perp,0) dv_\perp \bigl/\left( x_{m}y_{m}z_{m}\right)},
\end{eqnarray}
where $r_{o}$ and $r_{i}$ are outer and inner radii of the density inhomogeneity shown in Fig.~2 and $r_0$ is its
mid-point (centre).
$f_{e,i}$ are electron or ion velocity distribution functions. 
They are normalised in such a way that when integrated by all spatial and velocity
components they give the total number of real 
electrons and ions (not pseudo-particles that usually mimic much larger number of real particles
in the PIC simulation). The definition 
used in Eq.(7) effectively provides the fraction
(the percentage) of accelerated electrons, parallel to the magnetic field, in the density gradient regions.
Eq.(8) gives the fraction
(the percentage) of heated ions, perpendicular to the magnetic field.
In the case of 2.5D similar indexes have been defined in Ref.\cite{2011PhPl...18i2903T}.
We refer interested reader for an underlying discussion of these indexes in the former reference.
We gather from Fig.~5(a)--5(c) that electron acceleration is confined only to the
parallel to the background magnetic field direction. This can be evidenced
by appearance of the bumps in the $f_e(v_x)$ at $t=t_{end}$, which correspond to
the electron beams produced by the generated $E_x=E_\parallel$ in the density inhomogeneity. 
By comparing Fig.~5(a) (electrons) with Fig.~5(d) (ions)
we see that no ion beams are generated. Instead we see from Figs. 5(e) and 5(f) that ions
are heated in the perpendicular direction as the distribution function at  $t=t_{end}$
widens (whilst no ion beams are produced).
Note that such transverse ion heating of plasma has been studied in the past too
 \cite{2004JGRA..10904205C,2004ApJ...605L.149V,2006A&A...452L...7W,2007ApJ...659.1693W}, while
the main focus here is the electron acceleration in the parallel direction.
Comparing Fig.~5(a) (electrons) with Fig.~5(d) (ions) also gives us an indication that because
in the density gradient region (circular ring) ions essentially do not feel IAW driving in the
parallel direction while electrons are accelerated,  {\it effective charge separation} takes place.
This is consistent with the cold plasma approximation, see animation 3 (in Supplementary Data) 
from Ref.\cite{2007NJPh....9..262T}.
We  gather from Fig.~5(g) that the value of the generated parallel electric field, in normalised units,  is 0.07.
It can be also seen from Fig.5(h) that $AH_{\parallel,e}(t)$ and $AH_{\perp,i}(t)$ indexes
attain values of 0.45 (45\%) and 0.75 (75\%) respectively.
Fig.~3 from Ref.\cite{2011PhPl...18i2903T} directly corresponds to our Fig.~5 here but now for the 3D case.
Thus we can make some comparisons. The differences are:
(i) in 3D case parallel electron beams are more pronounced, as the solid line in Fig.~5(a) attains higher values also over the
larger $v_x$ range.
(ii) in 3D case the transverse ion heating is more pronounced too, as the solid lines in Figs.~5(e) and 5(f)
reach higher values compared to the 2.5D case.
(iii) in 3D case parallel electric field attains values roughly a factor of two larger,  as 
Fig.~5(g) shows that the value of the parallel electric field attains 0.07 versus 0.03 in the 2.5D case.
(iv) the fraction of accelerated electrons, parallel to the magnetic field, in the density inhomogeneity
is factor of two larger, as it can be seen from Fig.~5(h) that 
$AH_{\parallel,e}(t_{end})=0.45$ versus $AH_{\parallel,e}(t_{end})=0.2$ in the 2.5D case
and the fraction of heated ions, perpendicular to the magnetic field,
is also factor of two larger, as it can be seen from Fig.~5(h) that 
$AH_{\perp,i}(t_{end})=0.75$ versus $AH_{\perp,i}(t_{end})=0.35$ in the 2.5D case.

Fig.~6 presents the same data in panels (a)--(f) from Fig.~5 but now in terms of kinetic energy $E= (\gamma-1) m_e c^2$ and
also plotted on the log-log plot. Note that in Fig.~5 log-normal plot was used.
To produce Fig.~6 only $v_x>0$ branch was used. The reason for producing Fig.~6 is two-fold: 
(i) to demonstrate that the electron spectrum in the parallel to the field direction shows
so-called "knee" -- a break in the spectra power-laws;
(ii) to demonstrate that the "knee" can be interpreted is the Landau damping (Cerenkov resonance effect)
due to the wave-particle interactions.
There are several explanations for the observed double power-laws in the electron spectrum.
By double we mean one (usually smaller $\propto E^{-\alpha}$ with $2<\alpha<4$) power-law index below a certain fixed energy $E<E_{br}$ (where  50~keV~$< E_{br}<$~150~keV and
then another (usually larger 
$\propto E^{-\alpha}$ with $4<\alpha<7$) power-law index for  $E>E_{br}$ \cite{1992ApJ...389..756D}.
Chapter 13.3.4 of Ref.\cite{asch} quotes several explanations for the double power-law also known as the "knee".
The most common explanation is collisional relaxation of an electron beam, with added refinements such as the generation
of Langmuir waves and plasma density inhomogeneity \cite{2012A&A...539A..43K}.
Here we offer an alternative explanation in that flare generated DAWs produce electron beams on the transverse density inhomogeneities 
(magnetic loop edges) which in turn produce spectra with the structure that resembles the knee.

Fig.~7 is analogous to Fig.~6 but it has been produced with the simulation data from run L16 from Ref.\cite{2011PhPl...18i2903T} that corresponds to
2.5D. Comparison
of Fig.~7 (2.5D case, Ref.\cite{2011PhPl...18i2903T}) and Fig.~6 (3D case, this paper) enables to
grasp the differences between 2.5D and 3D geometry. It is evident from Figs.~7 and 6 that in the 3D case 
acceleration of electrons along the magnetic field and heating of ions in the transverse direction 
is more efficient. Also, the knee feature in the parallel to the magnetic field electron spectra can be seen.

Fig.~8 is analogous to Fig.~6 but  it has been produced with the simulation data from run L73 from Ref.\cite{2011PhPl...18i2903T} that corresponds to
2.5D. Comparison
of Fig.~8 (2.5D case, Ref.\cite{2011PhPl...18i2903T} mass ratio $m_i/m_e=73.44$), 
Fig.~7 (2.5D case, Ref.\cite{2011PhPl...18i2903T} mass ratio $m_i/m_e=16$) and  Fig.~6 (3D case, this paper, mass ratio $m_i/m_e=16$) enables to
gauge the differences between 2.5D versus 3D geometry as well as one mass ratio versus another (note that 
considered mass ratios land the physical system into different regimes. Case of $m_i/m_e=16$ corresponds to IAW regime, whereas $m_i/m_e=73.44$ 
corresponds to KAW regime). We gather from Fig.~8 that double power-law fit to the parallel electron spectrum is better
compared to Figs.~7 and 6, possibly indicating that the KAW regime provides more double power-law-like behaviour. It should be noted that in Fig.~7 and 6 
corresponding to IAW regime power-law fit seems somewhat arbitrary. This partly could also be due to the fact that
the observations usually provide time-averaged electron spectrum whereas no-time averaging has been used in Figs.6--8.
A more  profound result is that the energy, $E_{DAW}$, that corresponds to the break in the power-law, $E_{br}$,
{\it approximately} corresponds the energy of an electron which moves with {\it the phase speed of the DAW}, $V_{ph}$. 
In other words we note that as we increase the mass ratio from 16 to 73.44 both
the break in the spectrum and  $V_{ph}$ decrease. This demonstrates that 
 the "knee" can possibly be interpreted as the Landau damping (Cerenkov resonance effect) of DAWs
due to the wave-particle interactions.

\section{Conclusions}

The aim of this work was to explore the novelties brought about by 3D geometry effects into the 
problem of particle acceleration by DAWs in the solar flare and also to Earth magnetosphere 
auroral zone studied in an earlier 2.5D geometry work \cite{2011PhPl...18i2903T}.
It should be noted that here we have 
studied the case of plasma over-density, transverse to the background magnetic field, 
whereas references that deal with Earth auroral zone usually
consider the case of plasma under-density (a cavity), as 
dictated by the different applications considered (solar coronal loops and Earth 
auroral plasma cavities). As far as the generation of parallel electric field and
associated particle acceleration is concerned, there is no difference whether the transverse density gradient
is positive or negative -- for the mechanism to work it has to be non-zero.
This is because whistlers (and ion cyclotron waves) that propagate strictly along
the magnetic field display no Landau damping, since the longitudinal component of
the wave electric field is zero \cite{zsn}, page 124. The latter becomes non-zero as the wave front turns due to the transverse density
inhomogeneity. 
Thus we investigated a process of particle acceleration by left-hand, circularly polarised inertial Alfven wave
in a transversely inhomogeneous plasma, using 3D particle-in-cell simulation.
We considered a cylindrical tube that contains a transverse to the background magnetic field inhomogeneity
with a scale of the order of ion inertial length. 
Such numerical setup is relevant for solar flaring loops and
earth auroral zone.
In such structure IAWs with frequency $0.3 \omega_{ci}$ 
were launched  and allowed to develop three wavelength.
The following key points have been established:
Propagation of IAW in such a system generates
time-varying parallel electric field, localised in the density gradient
regions, which accelerate electrons in the parallel to magnetic field direction.
Perpendicular electric field of IAW also effectively heats ions in the transverse direction. 
The generated parallel electric field rotates in the same direction and frequency as the "parent" IAW.  
The fully 3D kinetic simulation demonstrates electron
acceleration efficiency in the density inhomogeneity regions, along the magnetic field, is of the order of 45\% and 
ion heating, in the transverse to the magnetic field direction,  is about 75\%.
The latter is a factor of two times higher than the previous 2.5D analogous study
and is broadly in agreement with the solar flare particle acceleration observations.
Log-log plots of electron spectra seem to indicate that 
the "knee", frequently seen in the solar flare observations, can 
be interpreted is the Landau damping of IAWs
due to the wave-particle interactions.

\begin{acknowledgments}
The author would like to thank EPSRC-funded 
Collaborative Computational Plasma Physics  (CCPP) project
lead by Prof. T.D. Arber (Warwick) for providing 
EPOCH Particle-in-Cell code. 
Computational facilities used are that of Astronomy Unit, 
Queen Mary University of London and STFC-funded UKMHD consortium at St Andrews
University. The author is financially supported by STFC consolidated grant ST/J001546/1, 
Leverhulme Trust research grant RPG-311, and HEFCE-funded 
South East Physics Network (SEPNET).
\end{acknowledgments}


\end{document}